\documentclass[manuscript, screen]{acmart}

\usepackage{todonotes}
\usepackage{subfig}
\AtBeginDocument{%
  }

\begin{document}

\title{Graphing the Everyday: A Neurosymbolic Approach to Eliciting Routines for Just-In-Time Adaptive Interventions}

%


\author{Shakyani Jayasiriwardene}
\authornote{Both authors contributed equally to this research.}
\authornote{Corresponding author.}
\affiliation{%
  \institution{The University of Sydney}
  \country{Australia}
}
\email{djay0399@uni.sydney.edu.au}

\author{Blake Mountford}
\authornotemark[1]
\affiliation{%
  \institution{The University of Sydney}
  \country{Australia}
}
\email{bmou8324@uni.sydney.edu.au}

\author{Meican Ma}
\affiliation{%
  \institution{The University of Sydney}
  \country{Australia}
  }
\email{mema2458@uni.sydney.edu.au}

\author{Niels van Berkel}
\affiliation{%
  \institution{Aalborg University}
  \country{Denmark}
  }
\email{nielsvanberkel@cs.aau.dk}

\author{Nicholas Koemel}
\affiliation{%
  \institution{Monash University}
  \country{Australia}
  }
\email{nicholas.koemel@monash.edu}

\author{Matthew Ahmadi}
\affiliation{%
  \institution{Monash University}
  \country{Australia}
  }
\email{matthew.ahmadi@monash.edu}

\author{Jorge Goncalves}
\affiliation{%
  \institution{University of Melbourne}
  \country{Australia}
  }
\email{jorge.goncalves@unimelb.edu.au}

\author{Emmanuel Stamatakis}
\affiliation{%
  \institution{Monash University}
  \country{Australia}
  }
\email{emmanuel.stamatakis@monash.edu}

\author{Zhanna Sarsenbayeva}
\affiliation{%
  \institution{The University of Sydney}
  \country{Australia}
  }
\email{zhanna.sarsenbayeva@sydney.edu.au}

\renewcommand{\shortauthors}{Jayasiriwardene \& Mountford et al.}

\begin{abstract}
Just-In-Time Adaptive Interventions (JITAIs) increasingly rely on conversational agents to elicit user routines, yet translating fluid human dialogue into rigid schedule data remains a significant challenge. We conducted a qualitative investigation of a neurosymbolic pipeline, combining Large Language Models (LLMs) with a Neo4j knowledge graph, to map unstructured verbal narratives into actionable interventions. Through human-centric evaluation using natural-language playbacks, we identified a critical ``mental-model gap,'' where the linear extraction of LLMs clashes with hierarchical, non-linear human storytelling, causing severe entity fragmentation. Furthermore, we articulate an ``ecological mismatch,'' demonstrating that algorithmic schedule availability frequently ignores the user's fluctuating psychological receptivity and physical energy levels. To resolve these tensions, we propose actionable design heuristics, including routine piggybacking, adaptive negotiation, and scalable transparency. Ultimately, these guidelines provide a foundational framework for evolving rigid schedule-trackers into empathetic, context-aware proactive agents capable of supporting long-term health behavior change.
\end{abstract}



\maketitle

\section{Introduction}




Modern lifestyles have become increasingly sedentary, posing significant challenges to long-term health and well-being. To combat this increased sedentary behaviour, digital health design has shifted towards proactive, real-time support methodologies like Just-In-Time Adaptive Interventions (JITAI)~\cite{Nahum2018, Haag2025, Henry2025, Hardeman2019, Muller2017}. These interventions aim to nudge users towards positive (health) behaviour change. JITAI focuses on capitalising on states of opportunity, targeting periods of high susceptibility to positive health behaviour change so that interventions are delivered exactly when they are most effective~\cite{Nahum2018}.

Currently, many of the JITAI systems rely on continuous sensor monitoring to trigger alerts. However, research still lacks understanding of how users' own knowledge about their routines and schedules can be incorporated into proactive and adaptive interventions. Furthermore, existing issues with JITAI systems indicate that further information is required to complement the existing methods used for adaptivity. These issues include the risk of user annoyance~\cite{Cha2020, Iqbal2005} and intervention fatigue~\cite{Nahum2018} due to ineffective timing. Rather than continuously monitoring and abruptly interrupting users, we propose a personalised approach that identifies users' schedules early on to help identify opportune moments to nudge them towards positive change. Identifying opportune times to interrupt can reduce the amount of disruption caused to the user~\cite{Iqbal2005}. By leveraging the conversational capabilities of Large Language Models (LLMs), opportune times can be inferred from a brief voice interaction about a user's daily routine. This conversational elicitation also addresses a critical factor in JITAI: support must only be provided when a person is truly receptive~\cite{Nahum2018}.


However, relying on natural conversation introduces a significant human-computer alignment challenge. Human narratives about daily routines are inherently unstructured and contextual. In contrast, proactive AI systems require rigid, deterministic data to function reliably. Therefore, relying on LLMs as data extractors may lead to misalignment between the user's lived experience and the system's internal model~\cite{Emsley2023, Salvagno2023, Maleki2024, Vela2022}. This introduces a challenge of translation and sense-making. We must understand how users conceptually map their unstructured routines, and where systematic friction occurs when an AI attempts to translate those human narratives into rigid computational structures~\cite{Zhou2025}.

To address this gap between users' lived experience and LLM's interpretation, our study investigates how users verbally conceptualise their daily lives, and how this unstructured conversational data can be effectively translated into a structured format using knowledge graphs. Further, we seek to understand the inherent friction points that emerge when mapping unstructured human narratives onto rigid digital data structures. We implemented a qualitative approach to evaluate the representational fidelity of our graph-based system. First, we conducted a member-checking analysis by generating state-machine diagrams and knowledge graphs from user transcripts and asking users for feedback. Second, we conducted a human-centric evaluation by presenting a summary generated from the knowledge graphs back to the user, capturing their subjective assessment of whether the AI's mental model matched their own lived experience.

The key findings of our study shows that while users found the proactive voice agent highly natural, a distinct ``mental-model gap'' exists between how humans narrate their daily lives and how algorithms linearly extract them. Although standard zero-shot extraction frequently resulted in discrepancies, participants actively and willingly sought to repair the system's mental model when provided with structural transparency. Ultimately, our results show that pairing conversational fluency with scalable transparency empowers users to intuitively correct structural misalignments. This helps in successfully mitigating the ecological mismatch between algorithmic schedule availability and true psychological receptivity.


The core contribution of this work is a rigorous empirical understanding of the structural and behavioural friction inherent in conversational health agents for JITAI. This provides a foundation for more adaptive, context-aware systems. Specifically, our contributions are threefold: (a) we establish how users' unstructured conversational data can be effectively translated into a persistent neurosymbolic architecture, acting as a structured mental model that the system can reliably map and probe for contextual information; (b) we empirically identify the critical friction points, the ``mental-model gap'', that emerge when forcing complex, non-linear human dialogue into rigid data structures; and (c) we propose concrete, actionable design heuristics for the development of future proactive agents for JITAI.







\section{Related Work}

This section reviews three bodies of literature that motivate our work. The first is the evolution of Just-in-Time Adaptive Interventions (JITAIs) and the centrality of receptivity. The second is the use of conversational agents in health and behaviour change. The third is the challenge of translating unstructured human narratives into structured representations for AI systems.

\subsection{Just-in-Time Adaptive Interventions (JITAIs) and Receptivity}

Over the past decade, digital behaviour change interventions have shifted substantially. They have moved from static, scheduled messaging toward dynamic systems that adapt to the user's evolving context~\cite{Nahum2018, Hardeman2019, Henry2025, Free2013}. Early mHealth interventions successfully extended the reach of behavioural support. However, they were repeatedly criticised for one-size-fits-all delivery, which produced notification fatigue and limited long-term engagement~\cite{Muller2017, Free2013}. Just-in-Time Adaptive Interventions (JITAIs) were proposed to address these shortcomings. They formalise the principle of providing the ``right type of support, at the right time, in the right amount''~\cite{Nahum2018}. Within this framework, tailoring variables govern when and how interventions are delivered~\cite{Nahum2018, Klasnja2019}. These variables are dynamic indicators of the user's internal state, environment, or readiness.

In practice, these tailoring variables are operationalised largely through continuous passive sensing on smartphones and wearables. Such systems draw on signals such as step counts, GPS, heart-rate variability, application usage, and accelerometry~\cite{Rabbi2015, Klasnja2019, Sarker2014, Lane2010}. HeartSteps, for example, delivers contextualised activity prompts based on time of day, weather, and location~\cite{Klasnja2019}. Sense2Stop instead uses physiological sensing to detect stress and pre-empt smoking relapse~\cite{Battalio2021}. Such systems showcase the promise of context-aware computing, but they suffer from two well-documented limitations. First, raw sensor signals correlate only loosely with users' psychological availability~\cite{Mehrotra2017, Pielot2014}. A sensor can register inactivity, but cannot infer whether that inactivity is restorative, deliberate, or transient. Second, continuous sensing imposes non-trivial battery, privacy, and computational costs~\cite{Sarker2014, Lane2010}. Its inferences are often too coarse to ground precise intervention timing.

A complementary line of research emphasises receptivity: a user's momentary willingness and ability to engage with an intervention~\cite{Nahum2018, Pielot2014, Kunzler2019}. Receptivity is analytically distinct from contextual availability, and it strongly shapes JITAI effectiveness. Mistimed notifications, even when contextually plausible, are a primary driver of annoyance and disengagement~\cite{Iqbal2005, Cha2020, Pielot2014, Edwards}. Iqbal and Bailey~\cite{Iqbal2005} showed that interruptions at task boundaries are markedly less disruptive than those embedded mid-task. This finding motivated an extensive literature on inferring opportune moments from contextual cues~\cite{Mehrotra2017, Kunzler2019, Pielot2017}. Such models can predict receptivity to a degree~\cite{Kunzler2019, Mishra}. Yet breakpoints derived purely from sensor data still miss the longer-horizon temporal structure that governs availability, such as recurring meetings, family commitments, and commute patterns~\cite{Kunzler2019, Mehrotra2017}. Recent work on proactive speech agents adds a further nuance: users' perceptions of when a voice agent should speak are highly sensitive to its perceived adaptivity, partner model, and conversational context~\cite{Edwards, Cha2020}.

To bridge this gap, hybrid systems combine passive sensing with light-touch self-report. SitCoach, for instance, integrates user-reported context with sensor signals to time prompts against prolonged sedentary periods~\cite{Dantzig2013}. Related work elicits brief in-situ surveys to refine intervention delivery~\cite{vanBerkel2017, Mishra}. Most recently, Haag et al.~\cite{Haag2025} used LLMs to issue JITAIs for physical activity in cardiac rehabilitation, showing both the promise and the brittleness of LLMs in real-time intervention generation. These approaches, however, remain fundamentally reactive. They query, sense, or generate at the very moment the user should be receiving support, imposing cognitive cost precisely when receptivity is fragile. We propose a complementary direction. Rather than inferring receptive moments on the fly, we elicit the user's own account of their day in advance, through a brief conversational onboarding. A first-person account encodes more than free time: it signals when a person expects to be willing, and not merely free, to act. We therefore do not propose a complete JITAI, but a receptivity-elicitation layer. It supplies downstream just-in-time systems, including LLM-based ones like Haag et al.'s~\cite{Haag2025}, with a structured, user-authored account of when they expect to be reachable.

\subsection{Conversational Agents for Health and Behaviour Change}

Conversational agents (CAs) have a long heritage in HCI as media for health support. They span text-based chatbots for mental wellbeing~\cite{Fitzpatrick2017, Inkster2018}, embodied relational agents for chronic disease management~\cite{Bickmore2010}, and voice-based or proactive speech agents~\cite{Edwards, Cha2020}. More recently, LLM-powered agents have been used for triage, coaching, and reflective dialogue~\cite{Singhal2023, Jo2023, Haag2025, Park2023}. Systematic reviews of CAs in healthcare report two consistent findings~\cite{Laranjo2018, TudorCar2020}: conversational interfaces elicit richer self-reported information than form-based instruments, and perceived empathy and rapport raise engagement over time. These properties make CAs a natural fit for the personal, narrative information that JITAIs depend on but struggle to obtain through sensing alone.

A particularly relevant strand of work uses CAs to collect Ecological Momentary Assessments (EMAs): repeated self-reports of behaviour, affect, and context in everyday life~\cite{Stone1994, Shiffman2008}. Compared with fixed-form instruments, conversational formats can lower the friction of self-report and elicit richer, in-the-moment accounts in the user's own words~\cite{Schroeder2018, Maharjan2022, Kocielnik2018}. Voice-based agents in particular have been used for the in-situ self-report of affect and well-being, in the mobile contexts where users actually live their lives~\cite{Maharjan2022}. Yet free-form responses pose a downstream challenge despite their narrative richness. Their unstructured nature makes the data difficult for later components of the system to act upon~\cite{Kocielnik2018, Schroeder2018}. This is doubly problematic for proactive systems. Their timing decisions require precise temporal anchors, such as clock times, durations, and recurrences, that conversational disclosures rarely supply in canonical form.

In the context of behaviour change, CAs have been deployed to deliver micro-interventions, support goal-setting, and enact accountability~\cite{Bickmore2010, Schroeder2018}. A growing body of work explores LLM-based agents as proactive coaches that respond to user-disclosed context~\cite{Jo2023, Haag2025}. Across these systems, however, conversation is treated as a delivery channel for interventions, not as a source of structured knowledge about the user. As a result, the availability information that surfaces during onboarding or check-in dialogues is rarely retained in a form that can shape later timing. Our work reframes this relationship. At onboarding, the agent's role is not to coach but to listen and structure. It translates a brief narrative disclosure about the user's day into a persistent, machine-actionable representation of when and where the user expects to be available.

\subsection{Structuring Human Narratives for AI Systems}

LLMs excel at fluent, naturalistic dialogue, but a growing body of work documents their limits as faithful extractors of structured knowledge. They are prone to hallucination, producing confident but ungrounded outputs~\cite{Emsley2023, Salvagno2023, Maleki2024}. They also exhibit systematic weaknesses in temporal reasoning~\cite{Wang2024, Chu2024}. Even state-of-the-art models struggle to maintain consistent temporal anchors across multi-turn dialogue, and frequently confuse duration, ordering, and recurrence~\cite{Wang2024}. These difficulties are not confined to time. Extracting entities and the relations between them from free text remains error-prone, even in dedicated information-extraction settings~\cite{Wei2024, Pan2024}. Both failure modes are acute when an LLM must convert a casual description of a day into a precise, connected schedule. Small inaccuracies there can propagate into systematically mistimed downstream interventions.

To compensate, recent work pairs LLMs with explicit symbolic structures that ground generation in verifiable entities and relations~\cite{Pan2024, Edge2024, Yasunaga2021}. This neuro-symbolic pairing of a generative model with a structured store is the design stance our system adopts. A particularly active line of work uses LLMs themselves as zero-shot information extractors, prompting the model to surface typed entities and relations directly from unstructured text~\cite{Wei2024}. Knowledge graphs (KGs), broadly construed as typed entity-relation representations, have long served as a substrate for personal context in ubiquitous computing~\cite{Balog2019}. Work on personal KGs argues that lifelogs, calendars, and reminders are better modelled as graphs of persistent entities than as flat text spans~\cite{Balog2019}. In parallel, the LLM community has converged on retrieval- and graph-augmented generation to ground outputs in long-term context~\cite{Lewis2020, Edge2024, Pan2024}. Memory-oriented systems make a similar point: MemGPT~\cite{Packer2023}, generative-agent architectures~\cite{Park2023}, and long-term conversational memory designs~\cite{Zhou2025} show that explicit, structured memory improves coherence over extended interactions. Our system adopts this perspective at a lightweight scale. It uses an LLM extractor to accumulate typed entities (events, locations, and free-time windows) and the relations between them as the conversation unfolds.

Despite this momentum, two gaps motivate our study. First, most entity-relation extraction pipelines are developed and evaluated on encyclopaedic or enterprise data, where ground truth is well defined and entities are stable~\cite{Pan2024, Edge2024, Wei2024}. Little work examines how they perform on a user's own narrative account of their day, in which ambiguity, omissions, and contradictions are intrinsic features rather than errors to be corrected~\cite{Zhou2025}. Second, such pipelines are typically evaluated with extractive metrics (precision, recall, F1) or downstream task accuracy~\cite{Zhou2025}. They are rarely judged by whether the resulting representation reflects how the users themselves conceive of their own day. We are not aware of a prior study that places the user as the arbiter of representational fidelity in this setting. Our contribution is therefore less the observation that extraction is imperfect than the method and setting in which we examine it. Our study interrogates the alignment between unstructured human disclosure and the structured representations on which proactive, conversation-driven interventions depend.

\section{System Architecture}

\begin{figure}[hbt]
    \centering
    \includegraphics[width=1\linewidth]{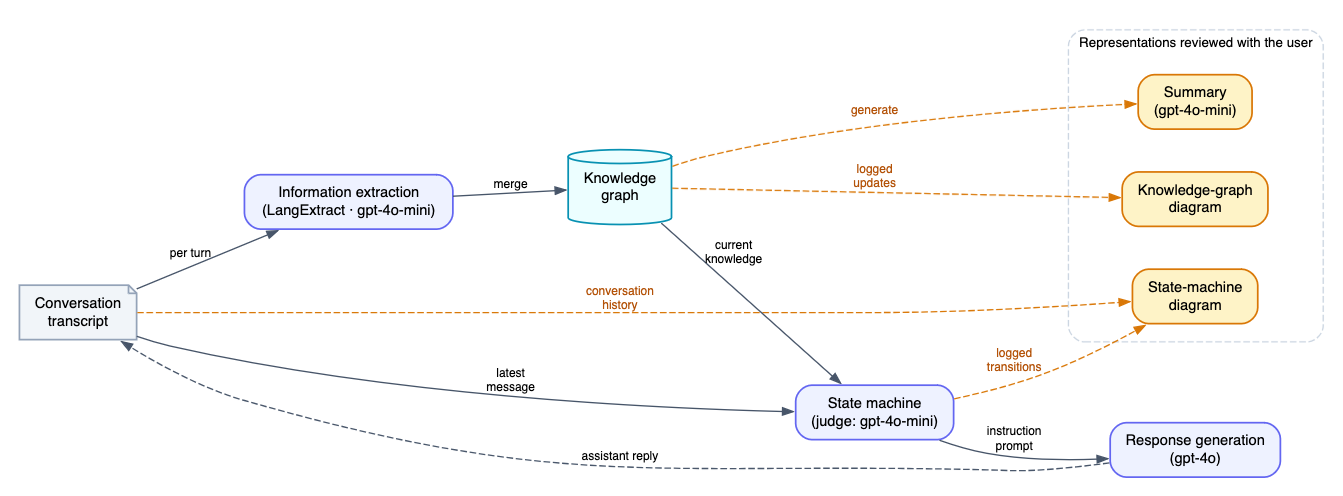}
    \caption{Schedule conversation system architecture. Each user utterance is extracted into a knowledge graph; the graph and latest message drive a state machine that generates each assistant reply. Logged transitions, graph updates, and the final graph produce the three representations reviewed with the user.}
    \label{fig:architecture2}
\end{figure}

Our system implements a receptivity-elicitation layer for an AI conversational agent, which identifies free time in the user's schedule by eliciting from the user an unstructured, conversational description of their day and then extracting structured event signals. Fig.~\ref{fig:architecture2} depicts the overall flow, from user utterance through extraction. As illustrated, the agent guides a short spoken discussion about the user's daily schedule, inquiring after details such as event times, their ordering, and commitments, and persists what it learns to a structured knowledge record. We adopt the neuro-symbolic design stance~\cite{Pan2024, Edge2024, Yasunaga2021}, pairing a generative language model with a knowledge graph that records the user's schedule as explicit entities and relations. At each turn (a dialogue between user and agent), the most recent transcript is processed by LangExtract \cite{Goel2025, Goel2026}, which extracts schedule entities (events, locations, and free-time windows) and the temporal relations between them. These extractions are merged into a personal knowledge graph \cite{Chakraborty2023, Hogan2021, Menschikov2026} that persists across daily conversations, letting the agent recall previously shared events and reason over the user's evolving routine \cite{Allan1983}.

    
    

\begin{figure}[hbt]
  \centering
  \subfloat[]{{\includegraphics[width=0.58\textwidth]{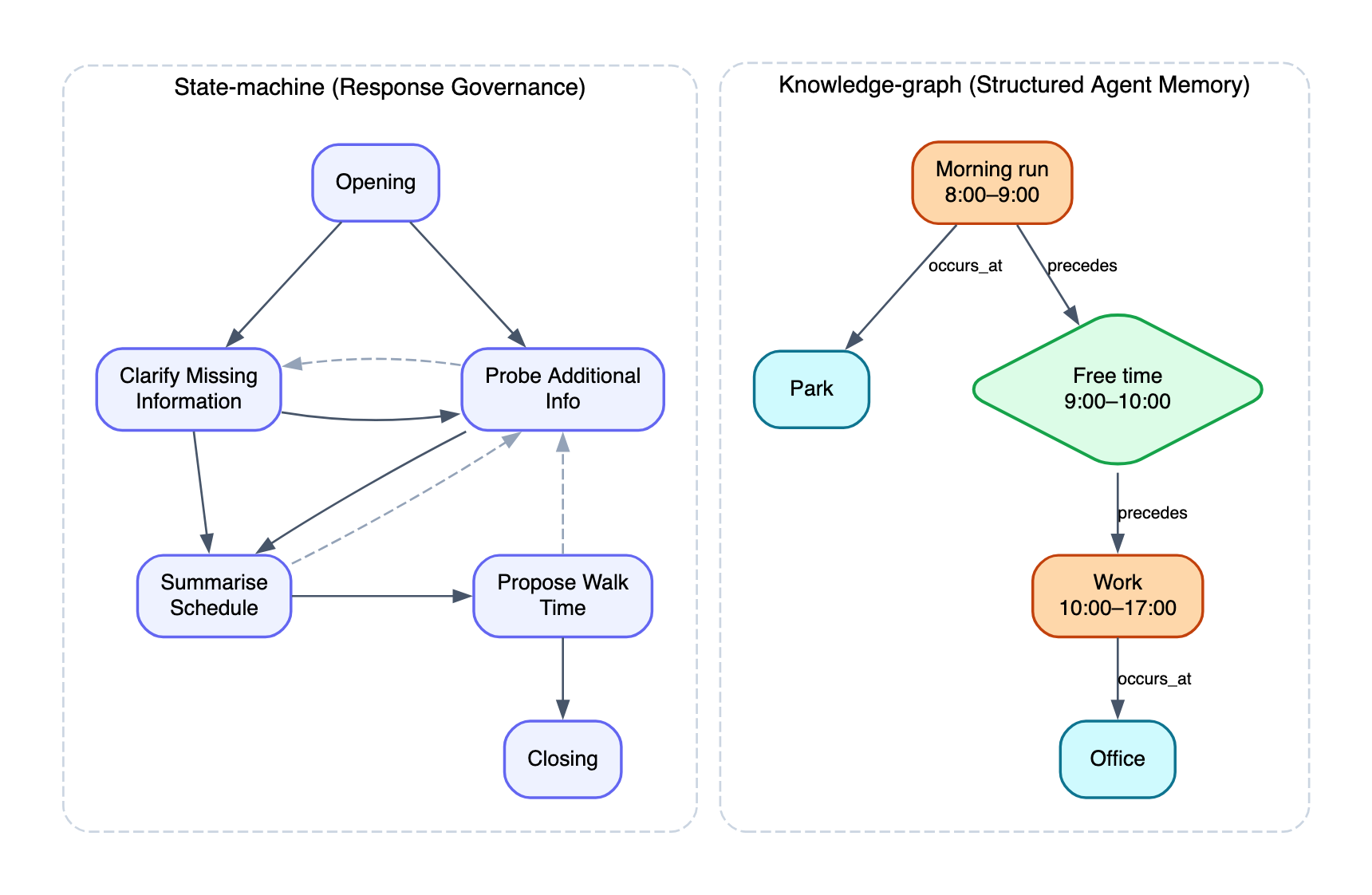} }\label{fig:sm_kg}}
  \hfill
  \subfloat[]{{\includegraphics[width=0.38\textwidth]{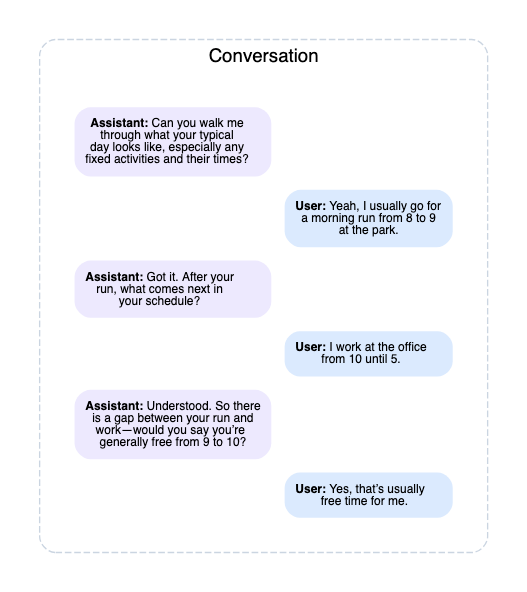} }\label{fig:transcript}}
  \caption{System mapping: (a) State-machine and knowledge-graph, alongside (b) corresponding conversation transcript.}
  \label{fig:sm_kg_transcript}
\end{figure}

Fig.~\ref{fig:sm_kg_transcript} depicts an example knowledge graph constructed from an unstructured conversation transcript, as well as the state machine~\cite{Zhu2010, Young2013, Kim2025, Rosen2026} that governs assistant response generation. Each state defines a conversational goal as a prompt template, with a set of transition rules that fire on either quantitative conditions (read directly from the knowledge graph) or semantic conditions (evaluated by a Judge-LLM subagent \cite{Zheng2023}). At every turn, the machine transitions on the current graph and the latest user message, a fresh instruction prompt is built from the active goal, and the assistant's reply is generated by OpenAI's \verb|gpt-4o| \cite{Openai2023}; extraction and the semantic judgements use \verb|gpt-4o-mini|. Together, the states and their transition rules form a conversational workflow in which the agent efficiently gathers the user's schedule, clarifies incomplete information, and confirms what it has understood before closing, returning to earlier states when the user revises what they have said.

\begin{figure}[hbt]
    \centering
    \includegraphics[width=1\linewidth]{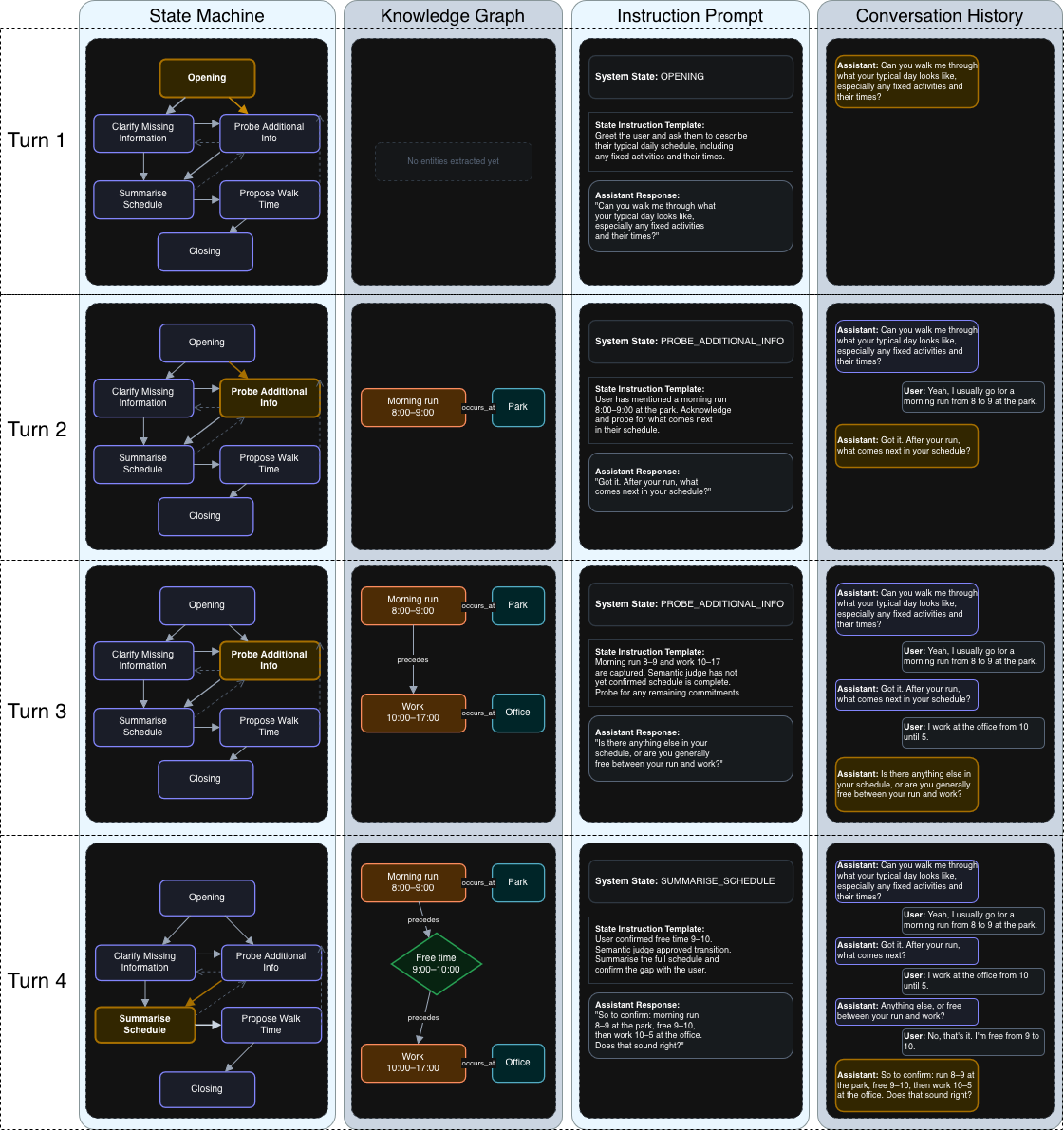}
    \caption{An example conversation, with the core system elements visualised at each turn, in a similar view to what the users are presented with during discussions. Each row shows one conversation turn: the state machine advances (amber), the knowledge graph accumulates entities from user utterances, and the instruction prompt is recomposed from both to guide the next response.}
    \label{fig:sm_kg_prompt_tran}
\end{figure}

 A significant benefit of structured governance is traceability \cite{Kim2026}: every response maps back to a discrete state and goal. Because the agent's memory and decision-making are structured as graphs, the representations it builds internally can be easily visualised and shown back to the user. After each conversation, we generated two trace diagrams from the system logs: the state-machine trace and knowledge-graph trace, represented in Fig.~\ref{fig:sm_kg_prompt_tran}. The state-machine trace displays the conversation transcript and a state machine diagram. It lets a user step through the conversation turn by turn, highlighting the active state and any activated transition rules, as well as the instruction prompt that produced the assistant response for that turn. The knowledge-graph trace shows the state of the knowledge graph at each point in the conversation: its events, locations, inferred free-time windows, as well as the relations between these entities. The knowledge graph trace also presents a short summary of the information contained within the final knowledge graph generated by \verb|gpt-4o-mini| for the user to check. Because these views make the system's reasoning visible, users could identify and correct the places where its model of their day differed from their own.

\section{Study Design}
\label{sec:study}
To evaluate the representational fidelity and human alignment of our graph-mapping pipeline, we conducted an in-the-lab evaluation study with 16 participants following HCI sample standards~\cite{Caine2016}. Given that the primary objective of this investigation is a deep qualitative analysis of human-AI translation friction, the selected sample size allows for the saturation of design insights without sacrificing the granularity required to analyse nuanced dialogue patterns. 

This study was approved by our Institutional Ethics Review Committee. Fig.~\ref{fig:methodology} illustrates the study flow, organised into a phased, semi-structured laboratory session designed to maximise participant comfort and minimise observation bias.:

\begin{figure}[H]
    \centering
    \includegraphics[width=0.85\linewidth]{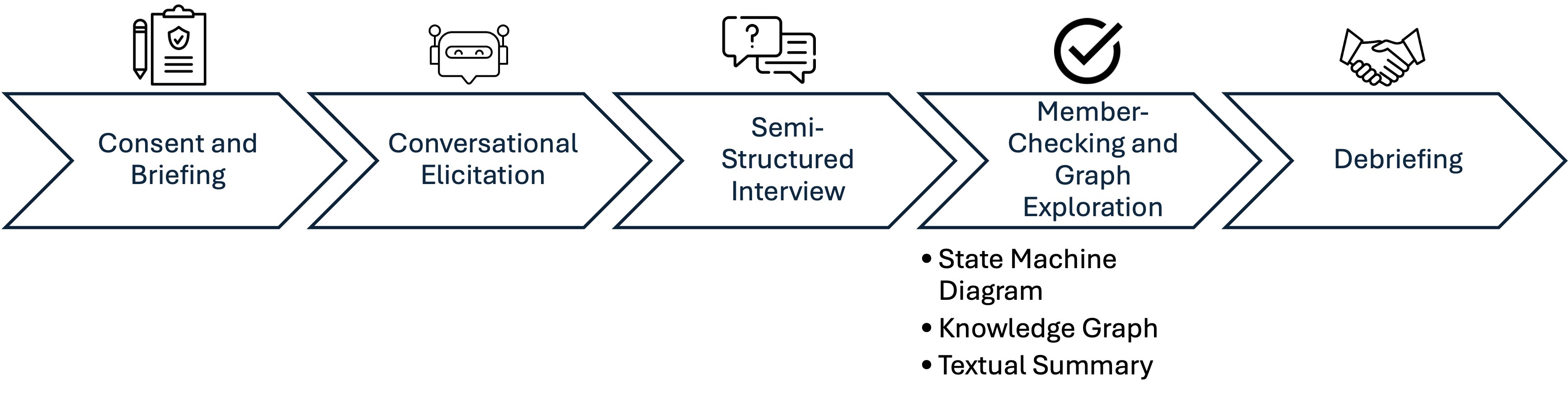}
    \caption{The study flow organised into five distinct phases: (1) Consent and Briefing (2) Conversational Elicitation (3) Semi-structured Interview (4) Member-Checking and Graph Exploration (5) Debriefing.}
    \label{fig:methodology}
\end{figure}

\begin{itemize}
    \item \textbf{Phase 1: Consent and Briefing.} Participants entered the research laboratory and were briefed on the objectives of the mobile application's conversational interface. Afterwards, they provided consent by signing the participant consent form. 
    \item \textbf{Phase 2: Conversational Elicitation.} To ensure naturalistic communication behaviour and prevent performance anxiety, researchers left the room during the interaction, allowing participants to comfortably outline their daily schedules directly to the voice assistant. 
    \item \textbf{Phase 3: Semi-Structured Interview.} Upon completion of the voice interaction, the researchers returned to the room to conduct a semi-structured contextual inquiry. Participants were initially asked for open-ended general feedback regarding their overall user experience.
    \item \textbf{Phase 4: Member-Checking and Graph Exploration.} To capture the structural translation friction of our data pipeline, researchers systematically introduced three distinct artifacts generated from the interaction: the state machine diagram, the Neo4j knowledge graph, and the textual summary through an interactive interface. Utilizing a concurrent Think-Aloud protocol, participants actively explored these structures and were prompted to audit the system for explicit mapping issues, identifying specific errors related to dropped context or structural sequencing.
    \item \textbf{Phase 5: Debriefing.} Once the interview and reviews were done, participants were debriefed and thanked for their participation.
\end{itemize}




\section{Qualitative Analysis}

\label{sec:qual}




Our qualitative data comprises the interview material gathered during the member-checking phase of the study. The sessions were transcribed, and participants' feedback was segmented into 221 codeable units, each a single evaluative statement. We analysed these data with codebook-based thematic analysis, a reliability-oriented variant of thematic analysis~\cite{Braun2006} in which a
structured, defined codebook is applied to the data and inter-rater reliability is reported~\cite{Boyatzis1998, Guest2012, MacQueen1998}. 

Two authors first developed an initial codebook of 27 codes, each with a definition, inclusion and exclusion criteria, and an anchor example. These codes were developed considering both the study's framing and a close reading of the data. To establish inter-coder reliability, both authors independently coded 20\% of the data units.  The initial Cohen's $\kappa$ of 0.32 indicated
that while the codes were conceptually grounded, several boundaries between codes required further refinement. The authors discussed each point of disagreement to resolve ambiguities in code definitions and inclusion/exclusion criteria through consensus. Following this discussion, both authors independetly coded a separate 10\% of data units, yielding a Cohen's $\kappa$  of 0.71, indicating substantial agreement~\cite{LandisKoch1977}. The second author then coded the remaining data, with the first author auditing a randomly selected 29\% to confirm that reliability was maintained ($\kappa = 0.79$). Finally, both authors collaboratively reviewed the full set of coded data to develop higher order themes, grouping the 27 codes in five themes reported below through an iterative process of comparison and discussion.

We report three themes. Theme 1 locates the gap between how participants narrated their  routines and how the system structured them. Theme 2 examines the conversational channel through which those narratives were elicited. Theme 3 shows how participants, supported by the visual artefacts, worked to close the gap. Frequencies are reported as the number of coded units (n).


\subsection{Theme 1: Human Algorithm Narrative Gap}
Participants narrated their days according to an internal logic that is coherent to humans but opaque to a linear extractor. The most immediate property was non-linear disclosure (n=9): information arrived out of order and reframed what came before. P1 traced an extraction error directly to this, noting that \textit{``the way I talked was a bit complicated, not in order, so the graph separates them''}, and P11 observed the system working against the grain of his speech: \textit{``it tries to organise [what I said] into a timeline, even if I said something random.''} A deeper articulation of the same problem was hierarchical disclosure (n=5). Participants narrated high-level activities that contain sub-events: P12 explained that coding, a meeting, and deployment were all part of one user study, yet the system \textit{``identified them as four independent things''}, and P1 concluded that the representation \textit{``should be hierarchical, not linear.''} The remaining properties show that the gap is multidimensional. Participants assumed the system would infer unstated context (n=7), as P14 admitted: \textit{``I forgot to say that I usually play games at uni. I assumed that the AI knows my schedule.''} Their time references resisted precise values (n=5), with P16 conceding \textit{``I don't exactly know what times I have to spend for walking between tasks''} and P13 describing \textit{``a pattern, not what I specifically do today.''} A few narrated conditionally linked events that the system treated as independent (n=3), as P2 anticipated: \textit{``sometimes there can be other dependencies; the state may not be the best way to map those.''} Together, these properties give human routine narratives an order, depth, and approximateness that a flat, turn-by-turn extraction does not expect.

These narrative properties mapped predictably onto the extraction failures participants identified. At the entity level (n=17), the system misrepresented what things are. Co-referent mentions split into duplicate nodes (n=12): the graph \textit{``didn't recognise the office and workplace are the same thing''} and stored \textit{``grocery store, grocery, supermarket''} as distinct places (P1), while P8 found \textit{``so many instances of work, while it is the same thing: working on the thesis.''} Continuous activity was also carved at the wrong grain (n=5), as when P5 found walking absorbed into another event: \textit{``some part of walking is embedded into heading home.''} At the event level (n=56), the system misrepresented when things happen: stated information went missing (n=20), order was wrong (n=20), and times were inaccurate (n=16). P2's dinner was absent \textit{``even though I mentioned it''}; P11's walk was placed before dinner instead of after; and stated commitments resurfaced as availability, with P4 reporting \textit{``I said I start study at 2pm, but it shows I have free time from 2--6pm''} and P15 finding that the system \textit{``identified non-free times as free times.''} Entity-type confusion (n=3) and speech recognition errors (n=3, e.g., \textit{``gym captured as dream''}, P6) appeared only as edge cases. Each failure type is thus the structural mirror of a narrative property: renaming produces duplicates, nesting produces fragmentation, and vagueness produces invented precision.

The gap also runs deeper than factual representation. Participants distinguished between being technically free and being ready to walk (n=4): P8 expected an existing routine to count, \textit{``I walk to my apartment from work, and I expected [it] to say we can fit the walk into that time''}, and others rejected technically valid windows (n=10), as P3 did \textit{``because of the time it selected and the energy levels I have.''} These psychological states are invisible to any knowledge graph and mark the limit of what extraction alone can capture. In relation to our research question, this theme shows that users conceptualise routines hierarchically, non-linearly, and approximately, and that each of these properties has a matching, systematic friction point in the translation to rigid structure.

\subsection{Theme 2: Conversational Experience}
The conversation itself was the strongest part of the experience. Perceived naturalness was the most evidenced code in the dataset (n=29), and it was actively noticed rather than merely not complained about: P8 judged that \textit{``overall, the conversation is quite natural''}, and P13 found it \textit{``straight to the point''}, with follow-up questions that \textit{``made sense.''} Beyond naturalness, participants experienced the exchange as collaborative (n=10). P9 felt the agent \textit{``wanted to work with me, not disruptive''} and \textit{``accepted my suggestions''}, while P11 described joint refinement: \textit{``it gave me directions and I tried to refine those directions.''} As a channel for eliciting routines, LLM-based voice interaction succeeded experientially in a way earlier rule-based systems could not.

The same agent that felt natural, however, was narrowly goal-driven (n=14), a limitation of the task structure beneath the LLM surface rather than of its fluency. P1 \textit{``felt strict: it wanted me to make a choice rather than having a conversation''}, and P14 \textit{``told it many times I don't need help from it, but it insists on giving a time.''} This task-narrowness also foreclosed more emotionally responsive engagement, with P3 wanting \textit{``emotional support and suggestions based on my walking experience''} rather than scheduling alone (n=2).

When participants pushed back, the system looped rather than adapted (n=12). Repetition was a symptom of this inflexibility, not an independent failure. P5, who had already given a fixed schedule, was asked the same question \textit{``about 3 times, which made me [feel like] talking to a person who doesn't understand what I'm telling [them]''}, and P10 noted that \textit{``humans don't ask things twice.''} The agent's inability to revise its path produced the feeling of not being listened to.

Reports that the conversation was too long (n=17) followed the same pattern. Length was an accumulation of insistence and repetition rather than a standalone parameter: participants who hit these loops wanted the interaction cut to \textit{``two or three minutes''} (P8), while those who did not judged the same protocol \textit{``not too long, perfect length''} (P9). For our research question, the conversational channel proved viable for eliciting unstructured routines, but its rigidity shaped both the experience and the narratives participants were able to give.
\subsection{Theme 3: Bridging the Gap}
Some participants worked to prevent misalignment before it occurred (n=4). They monitored the system's understanding in real time and volunteered unsolicited context: P4 \textit{``mentioned both 8pm at dinner and 8pm at gym to clarify that mistake''}, and P10 repeated key details twice to be safe. This is a form of metacognitive engagement: these users treated the AI's model of their day as something to be managed during the conversation itself.

The visual artefacts did not just display the system's model; they created the conditions for participants to perceive and articulate the gap. The state machine view opened the conversational logic to inspection (n=19). P5 used it to locate a failure between knowledge and behaviour, observing that \textit{``the instruction prompt looks like it captured the context better than what I got in the conversation''}, and P12 traced a conversational misstep to a single incorrect state transition. The knowledge graph opened the extracted content to verification (n=20). Participants audited it node by node, confirming \textit{``very accurate events, locations and relationships and order''} (P6) and catching errors they had not noticed while speaking; P15 contrasted this inspectability with opaque assistants, where \textit{``you don't really see what's happening in the background.''} Scaffolding therefore operated at two levels at once, exposing how the system reasons and what it believes.

What the scaffolding revealed, participants answered in two ways. Some absorbed the gap (n=13), treating residual errors as tolerable (P4), despite the errors documented in Theme~1. Others closed the gap cognitively (n=20), specifying what a better system would do: update and invalidate its own entities \textit{``even within the day''} (P1), learn \textit{``how my weekdays go, and how my Saturdays go, over a month''} (P5), and \textit{``ask you where you are having the event instead of assuming''} (P14). Participants acted as active evaluators rather than passive recipients, willing co-designers of the system's representation.

The positive response to post-hoc scaffolding generated a desire for it in real time (n=5). P2 wanted a live indication of \textit{``whether the information is enough to create a plan''}, and P11 wanted to \textit{``modify and directly edit''} the structure in the chat itself. These requests point toward a live, collaborative knowledge graph as a design direction, and toward visual scaffolding as a practical route to LLM transparency: users were willing to bridge the human-algorithm gap and asked for the tools to do so during the conversation, not after it.

\section{Discussion}


The findings of this study articulate a fundamental structural paradox at the intersection of natural language processing and ubiquitous health technologies. While contemporary Large Language Models (LLMs) excel at maintaining fluid, naturalistic onboarding dialogues, a distinct ``mental-model gap'' emerges when translating these fluid human narratives into the rigid, symbolic data structures required by proactive digital health systems~\cite{Zhou2025}. By exploring how users verbally map their unstructured daily routines, we have identified systematic translation friction points that offer critical design heuristics for future conversational Just-In-Time Adaptive Interventions (JITAIs)~\cite{Nahum2018}. We further suggest design guidelines that would help AI agents to understand their users for JITAIs better.

\subsection{The Mental-Model Gap: Hierarchical Narratives vs. Linear Extraction}

Our human-centric evaluation reveals that the primary challenge of graph-based routine elicitation is entity fragmentation and relational flattening. Humans conceptually structure their schedules top-down and hierarchically~\cite{Planer2023}; they speak in complex, non-linear ways, frequently injecting random mentions or minor details that they do not explicitly plan to include in a formal schedule. Standard zero-shot LLM extractors, conversely, operate bottom-up and linearly~\cite{Wang2024}.

This strict linear architecture causes severe node duplication and fragmentation. When systems fail to resolve co-references, such as failing to identify that ``workplace'' and ``office'', or ``grocery store'' and ``supermarket'', are identical entities, downstream granularity issues emerge~\cite{ReinhartzBerger2025}. These lead to missing context and incorrect chronological ordering. Furthermore, linear extraction often forces a single event to be split into unlinked sub-events. For example, ``heading home'' might be redundantly split into ``walking'' and ``taking the train'', without correctly indicating that they are both part of the same main event. To properly map human lives in alignment with natural cognitive event segmentation~\cite{Zacks2001}, states must be able to probe into hierarchical sub-events to capture a richer range of related information, including travel distance, duration, and the specific sequence of the itinerary~\cite{Farrahi2011}.

Finally, human narratives are characterised by unstated contexts where users simply forget to mention environmental dependencies~\cite{Clark1991}. As established in recent evaluations of conversational agents, time references expressed in natural dialogue rarely contain precise values, relying instead on temporal vagueness~\cite{CanabalJuanatey2024}. The extraction module must be equipped to handle this fuzziness appropriately to avoid invented precision, such as forcing an exact timestamp onto a casual timeframe. To prevent hallucinations, LLMs must be designed to actively ask questions to clarify ambiguity rather than automatically assuming constraints.

\subsection{Ecological Mismatch and Navigating User Boundaries}

A compelling tension emerged in our findings regarding algorithmic schedule availability versus true psychological receptivity. Having an empty time block mapped in a schedule does not guarantee a user has the energy or bandwidth to engage with an intervention~\cite{Nahum2018, Iqbal2005}. Contextual and psychological states, such as acute mental exhaustion or fluctuating energy levels, serve as strict boundaries to compliance. When proactive systems ignore these states, it triggers an ``ecological mismatch'', where the AI flawlessly parses the routine but entirely lacks the user's cognitive context. To prevent intervention fatigue, users indicated a preference for nudges that organically fit walks into their existing daily routines, such as extending a walking commute from the workplace to the apartment, rather than suggestions for isolated, high-friction workouts~\cite{Pinder2018}.

This ecological mismatch is severely exacerbated when the system lacks conversational adaptivity. While participants praised the natural fluency of the voice assistant, they experienced acute friction when the system's rigid, single-minded agenda collided with their real-world fatigue. Governed by a strict state machine, the agent occasionally yoked a highly fluent conversationalist to an immutable algorithmic goal. Consequently, when users pushed back, expressed exhaustion, or attempted to negotiate constraints, the system frequently looped aggressively, repeating itself to achieve its assigned data-extraction directive.

This inflexible behaviour shatters the illusion of shared context, instilling a feeling that the system is unempathetic and does not truly understand the user~\cite{Berube2024}. Ultimately, users prefer natural interactions enriched with emotional support and gentle, flexible suggestions over goal-driven force. To be effective, proactive speech agents must treat interactions as a bi-directional negotiation, displaying adaptivity and gracefully yielding their underlying agenda when users set boundaries.

\subsection{Co-Design, Scalable Transparency, and Neurosymbolic AI}

The findings of this study heavily validate the integration of Neurosymbolic AI in conversational health tools~\cite{Kim2026}. Purely LLM-driven voice coaches are fundamentally constrained by fixed context windows; over extended, multi-day sessions. They are highly prone to fragmented context degradation, temporal sequencing errors, and ``catastrophic forgetting''~\cite{Salwa2025}. By pairing the generative fluency of an LLM with the deterministic, structured memory of a Neo4j knowledge graph, the system is equipped to maintain verifiable, persistent context over time. Because the backend is symbolic, the system can systematically update itself, invalidating its own entities, merging duplicated nodes, or restructuring sub-events whenever a correction is issued.

Our evaluation demonstrated that this neurosymbolic architecture inherently affords a high degree of structural transparency. When participants were provided with a visual representation of the knowledge graph, revealing exactly what the AI ``thought'' about their routine, they readily transitioned from passive interactors to active, willing co-evaluators. This transparency helped users understand the AI's reasoning, directly advancing the principles of Explainable AI (XAI)~\cite{Dwivedi2023, Xu2019} and allowing researchers to precisely map the friction points of the algorithmic translation.

However, a critical design tension emerged regarding the cognitive load of this co-design process. Exposing the underlying conceptual model is an exceptionally powerful methodology for system evaluation, beta-testing, and explanatory debugging~\cite{Kulesza2015, Liao2020}. However, requiring everyday users to actively ``manage'' or manually correct an AI’s persistent memory in the wild is deeply impractical. It fundamentally contradicts the low-friction ethos of a Just-In-Time Adaptive Intervention (JITAI)~\cite{Nahum2018}.

Therefore, we advocate for scalable transparency in future neurosymbolic systems. For developers, researchers, and early adopters, exposing the structural graph allows for rapid debugging and trust calibration. For the end-user in a commercial deployment, however, the deterministic graph must remain in the background. The transparency of the neurosymbolic system should instead be abstracted into digestible, natural-language playbacks (e.g., ``It sounds like you usually walk to the station around 8 AM, did I get that right?''). This allows the agent to maintain the rigorous, self-correcting mental models required for long-term health behaviour change, while allowing the user to validate those models using low-friction voice commands rather than manual data management.

\subsection{Design Implications for Conversational JITAI Identification}

Translating our evaluation of the mental-model gap, ecological mismatch, and system rigidity into actionable HCI heuristics yields the following core design guidelines for future conversational health agents and Just-In-Time Adaptive Interventions:

\begin{itemize}
    \item \textbf{Design for Hierarchical Extraction:} Rather than forcing human dialogue into flat, linear schedules, underlying data structures must support hierarchical nesting. Systems must be capable of probing into overarching events (e.g., ``heading home'') and successfully breaking them down into rich sub-events (e.g., ``walking to the station,'' ``taking the train'') without duplicating nodes. Robust coreference resolution must be implemented to recognize synonymous entities (e.g., ``office'' and ``workplace'') to prevent context fragmentation~\cite{ReinhartzBerger2025, Farrahi2011}.

    \item \textbf{Clarification over Assumption:} Natural human dialogue is inherently vague regarding specific times and environmental dependencies~\cite{Clark1991}. Extraction modules must be programmed to accept ``fuzzy'' temporal bounds~\cite{CanabalJuanatey2024}. When critical context is unstated, the LLM should be prompted to ask clarifying questions rather than forcing an arbitrary, exact timestamp onto the knowledge graph, thereby preventing hallucinations and ``invented precision''~\cite{Wang2024}.

    \item \textbf{Contextual Anchoring:} To mitigate intervention fatigue and account for fluctuating psychological states, systems should avoid suggesting isolated, high-effort behavioural changes~\cite{Iqbal2005}. Instead, interventions should identify existing transitions in the user’s daily map and ``piggyback'' on them. For instance, suggesting a slightly longer walking route from the workplace to the apartment, aligning the nudge with the user's natural physical trajectory and current energy levels~\cite{Pinder2018}.

    \item \textbf{Adaptive Negotiation over Rigid Goal-Seeking:} Conversational agents must abandon strict, immutable state-machine agendas. If a user pushes back, attempts to negotiate a boundary, or expresses exhaustion, the system must recognise this friction immediately. The agent should temporarily pause its data-extraction or intervention goal and pivot to providing empathetic support, offering flexible, lower-friction alternatives rather than aggressively looping to achieve its original algorithmic directive~\cite{Berube2024}.

    \item \textbf{Design for Scalable Transparency:} To balance explainability with low cognitive load, designers must tailor how an agent's internal memory is exposed based on the user's role~\cite{Liao2020}. Direct interaction with the underlying neurosymbolic graph should be strictly reserved for developer debugging, trust calibration, and system evaluation~\cite{Kulesza2015}. For everyday deployment, this structural complexity must be abstracted from the end-user. Systems should instead employ conversational playbacks that seamlessly translate algorithmic assumptions into natural dialogue~\cite{Nahum2018}.
\end{itemize}

\subsection{Limitations and Future Work}
While this study provides a foundational understanding of the representational fidelity and structural friction inherent in neurosymbolic routine mapping, several limitations warrant future investigation. First, our evaluation was conducted in a controlled, in-the-lab setting with 16 participants. While this sample size proved sufficient for thematic saturation regarding translation friction and the ``mental-model gap'', it does not capture the high-variance, unpredictable nature of daily life where schedules are constantly interrupted by fluctuating workloads, acute stress, and physical fatigue.

Consequently, this work serves as the necessary technical precursor to a broader behavioural inquiry. Our future research will expand this framework into a longitudinal field deployment, utilizing an event-contingent Experience Sampling Method (ESM) to trigger real-time, proactive nudges over multi-week periods. This longitudinal extension will allow us to investigate the limits of algorithmic perfection, specifically, why users may reject or hyper-comply with system-generated reminders even when the underlying knowledge graph maintains high technical accuracy. By cross-referencing this self-reported user context with objective physiological markers (e.g., smartwatch activity data) and system action logs, we intend to develop a robust taxonomy of proactive AI rejection, formally mapping the boundaries where algorithmic schedule availability intersects with psychological receptivity.
\section{Conclusion}

This paper investigated the critical friction points that emerge when translating fluid human dialogue into the rigid data structures required by proactive digital health systems with Just-in-time adaptive interventions (JITAIs). Through a qualitative evaluation of a schedule-mapping conversational agent, we identified a fundamental ``mental-model gap'' where linear LLM extraction fails to capture the hierarchical and inherently vague nature of human routines, and an ``ecological mismatch'' where perfect algorithmic scheduling remains blind to fluctuating user energy levels and psychological receptivity. To resolve these tensions, we advocate for integrating Neurosymbolic AI into Just-In-Time Adaptive Interventions (JITAIs). By anchoring generative LLM fluency within the deterministic structure of a knowledge graph, systems can maintain persistent, verifiable context while offering scalable transparency, allowing users to effortlessly validate the AI’s mental model through natural-language playbacks rather than high-friction data management. Ultimately, by prioritizing routine piggybacking, adaptive negotiation, and scalable transparency, future proactive agents can evolve from rigid schedule-trackers into context-aware, empathetic partners capable of supporting long-term health behavior change.
\section{Acknowledgement}

The authors acknowledge the use of AI Tools (ChatGPT, Gemini, Claude) during the preparation of this manuscript to improve language clarity and readability. The tool was used solely for editing and proofreading purposes, and the final text was thoroughly reviewed and approved by all authors.


\bibliographystyle{ACM-Reference-Format}
\bibliography{ref}

@inproceedings{Edwards,
author = {Edwards, Justin and Doyle, Philip R. and Branigan, Holly P. and Cowan, Benjamin R.},
title = {Comparing Perceptions of Static and Adaptive Proactive Speech Agents},
year = {2024},
isbn = {9798400705113},
publisher = {Association for Computing Machinery},
address = {New York, NY, USA},
url = {https://doi.org/10.1145/3640794.3665548},
doi = {10.1145/3640794.3665548},
booktitle = {Proceedings of the 6th ACM Conference on Conversational User Interfaces},
articleno = {25},
numpages = {12},
location = {Luxembourg, Luxembourg},
series = {CUI '24}
}

@article{Mishra,
author = {Mishra, Varun and K\"{u}nzler, Florian and Kramer, Jan-Niklas and Fleisch, Elgar and Kowatsch, Tobias and Kotz, David},
title = {Detecting Receptivity for mHealth Interventions},
year = {2023},
issue_date = {June 2023},
publisher = {Association for Computing Machinery},
address = {New York, NY, USA},
volume = {27},
number = {2},
issn = {2375-0529},
url = {https://doi.org/10.1145/3614214.3614221},
doi = {10.1145/3614214.3614221},
journal = {GetMobile: Mobile Comp. and Comm.},
month = aug,
pages = {23–28},
numpages = {6}
}

@article{Nahum2018,
  title={Just-in-Time Adaptive Interventions (JITAIs) in Mobile Health: Key Components and Design Principles for Ongoing Health Behavior Support},
  author={Nahum-Shani, Inbal and Smith, Shawn N and Spring, Bonnie J and Collins, Linda M and Witkiewitz, Katie and Tewari, Ambuj and Murphy, Susan A},
  journal={Annals of Behavioral Medicine},
  volume={52},
  number={6},
  pages={446--462},
  year={2018},
  publisher={Oxford University Press},
  doi={10.1007/s12160-016-9830-8},
  issn={0883-6612}
}

@inproceedings{Haag2025,
author = {Haag, David and Kumar, Devender and Gruber, Sebastian and Hofer, Dominik P. and Sareban, Mahdi and Treff, Gunnar and Niebauer, Josef and Bull, Christopher N and Schmidt, Albrecht and Smeddinck, Jan David},
title = {The Last JITAI? Exploring Large Language Models for Issuing Just-in-Time Adaptive Interventions: Fostering Physical Activity in a Prospective Cardiac Rehabilitation Setting},
year = {2025},
isbn = {9798400713941},
publisher = {Association for Computing Machinery},
address = {New York, NY, USA},
url = {https://doi.org/10.1145/3706598.3713307},
doi = {10.1145/3706598.3713307},
booktitle = {Proceedings of the 2025 CHI Conference on Human Factors in Computing Systems},
articleno = {644},
numpages = {18},
location = {
},
series = {CHI '25}
}

@article{Henry2025,
  author   = {Henry, Lauren M. and Blay-Tofey, Morkeh and Haeffner, Clara E. and Raymond, Cassandra N. and Tandilashvili, Elizabeth and Terry, Nancy and Kiderman, Miryam and Metcalf, Olivia and Brotman, Melissa A. and Lopez-Guzman, Silvia},
  title    = {Just-In-Time Adaptive Interventions to Promote Behavioral Health: Protocol for a Systematic Review},
  journal  = {JMIR Research Protocols},
  year     = {2025},
  month    = {Feb},
  day      = {11},
  volume   = {14},
  pages    = {e58917},
  doi      = {10.2196/58917},
  issn     = {1929-0748},
  url      = {https://www.researchprotocols.org/2025/1/e58917}
}

@article{Hardeman2019,
  title={A systematic review of just-in-time adaptive interventions (JITAIs) to promote physical activity},
  author={Hardeman, Wendy and Houghton, Jane and Lane, Katie and Beeke, Anne and {Vase} and Lydia and {Johnston} and Marie and {Sutton} and Stephen},
  journal={International Journal of Behavioral Nutrition and Physical Activity},
  volume={16},
  number={1},
  pages={31},
  year={2019},
  publisher={BioMed Central},
  doi={10.1186/s12966-019-0792-7},
  issn={1479-5868}
}

@article{Muller2017,
  title={The conceptualization of a Just-In-Time Adaptive Intervention (JITAI) for the reduction of sedentary behavior in older adults},
  author={M{\"u}ller, Andre Matthias and Blandford, Ann and Yardley, Lucy},
  journal={MHealth},
  volume={3},
  pages={37},
  year={2017},
  month={Sep},
  publisher={AME Publishing Company},
  doi={10.21037/mhealth.2017.08.05},
  pmid={29184889},
  pmcid={PMC5682389},
  issn={2306-9740}
}

@article{Cha2020,
author = {Cha, Narae and Kim, Auk and Park, Cheul Young and Kang, Soowon and Park, Minkyu and Lee, Jae-Gil and Lee, Sangsu and Lee, Uichin},
title = {Hello There! Is Now a Good Time to Talk? Opportune Moments for Proactive Interactions with Smart Speakers},
year = {2020},
issue_date = {September 2020},
publisher = {Association for Computing Machinery},
address = {New York, NY, USA},
volume = {4},
number = {3},
url = {https://doi.org/10.1145/3411810},
doi = {10.1145/3411810},
journal = {Proc. ACM Interact. Mob. Wearable Ubiquitous Technol.},
month = sep,
articleno = {74},
numpages = {28}
}

@inproceedings{Iqbal2005,
author = {Iqbal, Shamsi T. and Bailey, Brian P.},
title = {Investigating the effectiveness of mental workload as a predictor of opportune moments for interruption},
year = {2005},
isbn = {1595930027},
publisher = {Association for Computing Machinery},
address = {New York, NY, USA},
url = {https://doi.org/10.1145/1056808.1056948},
doi = {10.1145/1056808.1056948},
booktitle = {CHI '05 Extended Abstracts on Human Factors in Computing Systems},
pages = {1489–1492},
numpages = {4},
location = {Portland, OR, USA},
series = {CHI EA '05}
}

@article{Emsley2023,
  title={ChatGPT: these are not hallucinations – they’re fabrications and falsifications},
  author={Emsley, Robin},
  journal={Schizophrenia},
  volume={9},
  number={1},
  pages={52},
  year={2023},
  publisher={Nature Publishing Group UK},
  doi={10.1038/s41537-023-00379-4},
  issn={2334-265X}
}

@article{Salvagno2023,
  title={Artificial intelligence hallucinations},
  author={Salvagno, Michele and Taccone, Fabio Silvio and Gerli, Alberto Giovanni},
  journal={Critical Care},
  volume={27},
  number={1},
  pages={180},
  year={2023},
  publisher={BioMed Central},
  doi={10.1186/s13054-023-04473-y},
  issn={1364-8535}
}

@INPROCEEDINGS{Maleki2024,
  author={Maleki, Negar and Padmanabhan, Balaji and Dutta, Kaushik},
  booktitle={2024 IEEE Conference on Artificial Intelligence (CAI)}, 
  title={AI Hallucinations: A Misnomer Worth Clarifying}, 
  year={2024},
  volume={},
  number={},
  pages={133-138},
  doi={10.1109/CAI59869.2024.00033}
  }

@article{Vela2022,
  title={Temporal quality degradation in AI models},
  author={Vela, David and Sharp, Andrew and Zhang, Ruizhe and others},
  journal={Scientific Reports},
  volume={12},
  number={1},
  pages={11654},
  year={2022},
  publisher={Nature Publishing Group UK},
  doi={10.1038/s41598-022-15245-z},
  issn={2045-2322}
}

@misc{Zhou2025,
      title={A Simple Yet Strong Baseline for Long-Term Conversational Memory of LLM Agents}, 
      author={Sizhe Zhou and Jiawei Han},
      year={2025},
      eprint={2511.17208},
      archivePrefix={arXiv},
      primaryClass={cs.CL},
      url={https://arxiv.org/abs/2511.17208}, 
}

@misc{Wei2024,
      title={ChatIE: Zero-Shot Information Extraction via Chatting with ChatGPT}, 
      author={Xiang Wei and Xingyu Cui and Ning Cheng and Xiaobin Wang and Xin Zhang and Shen Huang and Pengjun Xie and Jinan Xu and Yufeng Chen and Meishan Zhang and Yong Jiang and Wenjuan Han},
      year={2024},
      eprint={2302.10205},
      archivePrefix={arXiv},
      primaryClass={cs.CL},
      url={https://arxiv.org/abs/2302.10205}, 
}

@article{Klasnja2019,
  author    = {Klasnja, Predrag and Smith, Shawna and Seewald, Nicholas J. and Lee, Andy and Hall, Kelly and Luers, Brook and Hekler, Eric B. and Murphy, Susan A.},
  title     = {Efficacy of Contextually Tailored Suggestions for Physical Activity: A Micro-randomized Optimization Trial of HeartSteps},
  journal   = {Annals of Behavioral Medicine},
  volume    = {53},
  number    = {6},
  pages     = {573--582},
  year      = {2019},
  publisher = {Oxford University Press},
  doi       = {10.1093/abm/kay067},
  issn      = {0883-6612}
}

@article{Free2013,
  author    = {Free, Caroline and Phillips, Gemma and Galli, Leandro and Watson, Louise and Felix, Lambert and Edwards, Phil and Patel, Vikram and Haines, Andy},
  title     = {The Effectiveness of Mobile-Health Technology-Based Health Behaviour Change or Disease Management Interventions for Health Care Consumers: A Systematic Review},
  journal   = {PLOS Medicine},
  volume    = {10},
  number    = {1},
  pages     = {e1001362},
  year      = {2013},
  publisher = {Public Library of Science},
  doi       = {10.1371/journal.pmed.1001362},
  issn      = {1549-1676}
}

@article{Lane2010,
  author    = {Lane, Nicholas D. and Miluzzo, Emiliano and Lu, Hong and Peebles, Daniel and Choudhury, Tanzeem and Campbell, Andrew T.},
  title     = {A Survey of Mobile Phone Sensing},
  journal   = {IEEE Communications Magazine},
  volume    = {48},
  number    = {9},
  pages     = {140--150},
  year      = {2010},
  publisher = {IEEE},
  doi       = {10.1109/MCOM.2010.5560598},
  issn      = {0163-6804}
}

@inproceedings{Sarker2014,
  author    = {Sarker, Hillol and Sharmin, Moushumi and Ali, Amin Ahsan and Rahman, Md. Mahbubur and Bari, Rummana and Hossain, Syed Monowar and Kumar, Santosh},
  title     = {Assessing the Availability of Users to Engage in Just-in-Time Intervention in the Natural Environment},
  year      = {2014},
  isbn      = {9781450329682},
  publisher = {Association for Computing Machinery},
  address   = {New York, NY, USA},
  url       = {https://doi.org/10.1145/2632048.2636082},
  doi       = {10.1145/2632048.2636082},
  booktitle = {Proceedings of the 2014 ACM International Joint Conference on Pervasive and Ubiquitous Computing},
  pages     = {909--920},
  numpages  = {12},
  location  = {Seattle, Washington},
  series    = {UbiComp '14}
}

@inproceedings{Pielot2014,
  author    = {Pielot, Martin and de Oliveira, Rodrigo and Kwak, Haewoon and Oliver, Nuria},
  title     = {Didn't You See My Message? Predicting Attentiveness to Mobile Instant Messages},
  year      = {2014},
  isbn      = {9781450324731},
  publisher = {Association for Computing Machinery},
  address   = {New York, NY, USA},
  url       = {https://doi.org/10.1145/2556288.2556973},
  doi       = {10.1145/2556288.2556973},
  booktitle = {Proceedings of the SIGCHI Conference on Human Factors in Computing Systems},
  pages     = {3319--3328},
  numpages  = {10},
  location  = {Toronto, Ontario, Canada},
  series    = {CHI '14}
}

@article{Pielot2017,
  author    = {Pielot, Martin and Cardoso, Bruno and Katevas, Kleomenis and Serr{\`a}, Joan and Matic, Aleksandar and Oliver, Nuria},
  title     = {Beyond Interruptibility: Predicting Opportune Moments to Engage Mobile Phone Users},
  journal   = {Proceedings of the ACM on Interactive, Mobile, Wearable and Ubiquitous Technologies},
  volume    = {1},
  number    = {3},
  articleno = {91},
  numpages  = {25},
  year      = {2017},
  month     = sep,
  publisher = {Association for Computing Machinery},
  address   = {New York, NY, USA},
  doi       = {10.1145/3130956},
  issn      = {2474-9567}
}

@article{Kunzler2019,
  author    = {K\"{u}nzler, Florian and Mishra, Varun and Kramer, Jan-Niklas and Kotz, David and Fleisch, Elgar and Kowatsch, Tobias},
  title     = {Exploring the State-of-Receptivity for mHealth Interventions},
  journal   = {Proceedings of the ACM on Interactive, Mobile, Wearable and Ubiquitous Technologies},
  volume    = {3},
  number    = {4},
  articleno = {140},
  numpages  = {27},
  year      = {2019},
  month     = dec,
  publisher = {Association for Computing Machinery},
  address   = {New York, NY, USA},
  doi       = {10.1145/3369805},
  issn      = {2474-9567}
}

@article{Mehrotra2017,
  author    = {Mehrotra, Abhinav and Pejovic, Veljko and Vermeulen, Jo and Hendley, Robert and Musolesi, Mirco},
  title     = {My Phone and Me: Understanding People's Receptivity to Mobile Notifications},
  year      = {2016},
  isbn      = {9781450333627},
  publisher = {Association for Computing Machinery},
  address   = {New York, NY, USA},
  url       = {https://doi.org/10.1145/2858036.2858566},
  doi       = {10.1145/2858036.2858566},
  booktitle = {Proceedings of the 2016 CHI Conference on Human Factors in Computing Systems},
  pages     = {1021--1032},
  numpages  = {12},
  location  = {San Jose, California, USA},
  series    = {CHI '16}
}

@article{Battalio2021,
  author    = {Battalio, Samuel L. and Conroy, David E. and Dempsey, Walter and Liao, Peng and Menictas, Marianne and Murphy, Susan and Nahum-Shani, Inbal and Qian, Tianchen and Kumar, Santosh and Spring, Bonnie},
  title     = {Sense2Stop: A Micro-randomized Trial Using Wearable Sensors to Optimize a Just-in-Time-Adaptive Stress Management Intervention for Smoking Relapse Prevention},
  journal   = {Contemporary Clinical Trials},
  volume    = {109},
  pages     = {106534},
  year      = {2021},
  publisher = {Elsevier},
  doi       = {10.1016/j.cct.2021.106534},
  issn      = {1551-7144}
}

@inproceedings{Rabbi2015,
  author    = {Rabbi, Mashfiqui and Aung, Min Hane and Zhang, Mi and Choudhury, Tanzeem},
  title     = {MyBehavior: Automatic Personalized Health Feedback from User Behaviors and Preferences Using Smartphones},
  year      = {2015},
  isbn      = {9781450335744},
  publisher = {Association for Computing Machinery},
  address   = {New York, NY, USA},
  url       = {https://doi.org/10.1145/2750858.2805840},
  doi       = {10.1145/2750858.2805840},
  booktitle = {Proceedings of the 2015 ACM International Joint Conference on Pervasive and Ubiquitous Computing},
  pages     = {707--718},
  numpages  = {12},
  location  = {Osaka, Japan},
  series    = {UbiComp '15}
}

@article{vanBerkel2017,
  author    = {van Berkel, Niels and Ferreira, Denzil and Kostakos, Vassilis},
  title     = {The Experience Sampling Method on Mobile Devices},
  journal   = {ACM Computing Surveys},
  volume    = {50},
  number    = {6},
  articleno = {93},
  numpages  = {40},
  year      = {2017},
  month     = dec,
  publisher = {Association for Computing Machinery},
  address   = {New York, NY, USA},
  doi       = {10.1145/3123988},
  issn      = {0360-0300}
}

@article{Fitzpatrick2017,
  author    = {Fitzpatrick, Kathleen Kara and Darcy, Alison and Vierhile, Molly},
  title     = {Delivering Cognitive Behavior Therapy to Young Adults With Symptoms of Depression and Anxiety Using a Fully Automated Conversational Agent ({Woebot}): A Randomized Controlled Trial},
  journal   = {JMIR Mental Health},
  volume    = {4},
  number    = {2},
  pages     = {e19},
  year      = {2017},
  publisher = {JMIR Publications Inc.},
  doi       = {10.2196/mental.7785},
  issn      = {2368-7959}
}

@article{Inkster2018,
  author    = {Inkster, Becky and Sarda, Shubhankar and Subramanian, Vinod},
  title     = {An Empathy-Driven, Conversational Artificial Intelligence Agent ({Wysa}) for Digital Mental Well-Being: Real-World Data Evaluation Mixed-Methods Study},
  journal   = {JMIR mHealth and uHealth},
  volume    = {6},
  number    = {11},
  pages     = {e12106},
  year      = {2018},
  publisher = {JMIR Publications Inc.},
  doi       = {10.2196/12106},
  issn      = {2291-5222}
}

@article{Bickmore2010,
  author    = {Bickmore, Timothy W. and Mitchell, Suzanne E. and Jack, Brian W. and Paasche-Orlow, Michael K. and Pfeifer, Laura M. and O'Donnell, Julie},
  title     = {Response to a Relational Agent by Hospital Patients with Depressive Symptoms},
  journal   = {Interacting with Computers},
  volume    = {22},
  number    = {4},
  pages     = {289--298},
  year      = {2010},
  publisher = {Oxford University Press},
  doi       = {10.1016/j.intcom.2009.12.001}
}

@article{Laranjo2018,
  author    = {Laranjo, Liliana and Dunn, Adam G. and Tong, Huong Ly and Kocaballi, Ahmet Baki and Chen, Jessica and Bashir, Rabia and Surian, Didi and Gallego, Blanca and Magrabi, Farah and Lau, Annie Y. S. and Coiera, Enrico},
  title     = {Conversational Agents in Healthcare: A Systematic Review},
  journal   = {Journal of the American Medical Informatics Association},
  volume    = {25},
  number    = {9},
  pages     = {1248--1258},
  year      = {2018},
  publisher = {Oxford University Press},
  doi       = {10.1093/jamia/ocy072},
  issn      = {1067-5027}
}

@article{TudorCar2020,
  author    = {Tudor Car, Lorainne and Dhinagaran, Dhakshenya Ardhithy and Kyaw, Bhone Myint and Kowatsch, Tobias and Joty, Shafiq and Theng, Yin-Leng and Atun, Rifat},
  title     = {Conversational Agents in Health Care: Scoping Review and Conceptual Analysis},
  journal   = {Journal of Medical Internet Research},
  volume    = {22},
  number    = {8},
  pages     = {e17158},
  year      = {2020},
  publisher = {JMIR Publications Inc.},
  doi       = {10.2196/17158},
  issn      = {1438-8871}
}

@article{Singhal2023,
  author    = {Singhal, Karan and Azizi, Shekoofeh and Tu, Tao and Mahdavi, S. Sara and Wei, Jason and Chung, Hyung Won and Scales, Nathan and Tanwani, Ajay and Cole-Lewis, Heather and Pfohl, Stephen and Payne, Perry and Seneviratne, Martin and Gamble, Paul and Kelly, Chris and Babiker, Abubakr and Sch{\"a}rli, Nathanael and Chowdhery, Aakanksha and Mansfield, Philip and Demner-Fushman, Dina and Ag{\"u}era y Arcas, Blaise and Webster, Dale and Corrado, Greg S. and Matias, Yossi and Chou, Katherine and Gottweis, Juraj and Tomasev, Nenad and Liu, Yun and Rajkomar, Alvin and Barral, Joelle and Semturs, Christopher and Karthikesalingam, Alan and Natarajan, Vivek},
  title     = {Large Language Models Encode Clinical Knowledge},
  journal   = {Nature},
  volume    = {620},
  number    = {7972},
  pages     = {172--180},
  year      = {2023},
  publisher = {Nature Publishing Group UK},
  doi       = {10.1038/s41586-023-06291-2},
  issn      = {1476-4687}
}

@inproceedings{Jo2023,
  author    = {Jo, Eunkyung and Epstein, Daniel A. and Jung, Hyunhoon and Kim, Young-Ho},
  title     = {Understanding the Benefits and Challenges of Deploying Conversational AI Leveraging Large Language Models for Public Health Intervention},
  year      = {2023},
  isbn      = {9781450394215},
  publisher = {Association for Computing Machinery},
  address   = {New York, NY, USA},
  url       = {https://doi.org/10.1145/3544548.3581503},
  doi       = {10.1145/3544548.3581503},
  booktitle = {Proceedings of the 2023 CHI Conference on Human Factors in Computing Systems},
  articleno = {18},
  numpages  = {16},
  location  = {Hamburg, Germany},
  series    = {CHI '23}
}

@article{Stone1994,
  author    = {Stone, Arthur A. and Shiffman, Saul},
  title     = {Ecological Momentary Assessment ({EMA}) in Behavioral Medicine},
  journal   = {Annals of Behavioral Medicine},
  volume    = {16},
  number    = {3},
  pages     = {199--202},
  year      = {1994},
  publisher = {Oxford University Press},
  doi       = {10.1093/abm/16.3.199},
  issn      = {0883-6612}
}

@article{Shiffman2008,
  author    = {Shiffman, Saul and Stone, Arthur A. and Hufford, Michael R.},
  title     = {Ecological Momentary Assessment},
  journal   = {Annual Review of Clinical Psychology},
  volume    = {4},
  pages     = {1--32},
  year      = {2008},
  publisher = {Annual Reviews},
  doi       = {10.1146/annurev.clinpsy.3.022806.091415},
  issn      = {1548-5943}
}

@inproceedings{Schroeder2018,
  author    = {Schroeder, Jessica and Wilkes, Chelsey and Rowan, Kael and Toledo, Arturo and Paradiso, Ann and Czerwinski, Mary and Mark, Gloria and Linehan, Marsha M.},
  title     = {Pocket Skills: A Conversational Mobile Web App To Support Dialectical Behavioral Therapy},
  year      = {2018},
  isbn      = {9781450356206},
  publisher = {Association for Computing Machinery},
  address   = {New York, NY, USA},
  url       = {https://doi.org/10.1145/3173574.3173972},
  doi       = {10.1145/3173574.3173972},
  booktitle = {Proceedings of the 2018 CHI Conference on Human Factors in Computing Systems},
  articleno = {398},
  numpages  = {15},
  location  = {Montreal QC, Canada},
  series    = {CHI '18}
}

@article{Kocielnik2018,
  author    = {Kocielnik, Rafal and Xiao, Lillian and Avrahami, Daniel and Hsieh, Gary},
  title     = {Reflection Companion: A Conversational System for Engaging Users in Reflection on Physical Activity},
  journal   = {Proceedings of the ACM on Interactive, Mobile, Wearable and Ubiquitous Technologies},
  volume    = {2},
  number    = {2},
  articleno = {70},
  numpages  = {26},
  year      = {2018},
  month     = jul,
  publisher = {Association for Computing Machinery},
  address   = {New York, NY, USA},
  doi       = {10.1145/3214273},
  issn      = {2474-9567}
}

@inproceedings{Park2023,
  author    = {Park, Joon Sung and O'Brien, Joseph and Cai, Carrie Jun and Morris, Meredith Ringel and Liang, Percy and Bernstein, Michael S.},
  title     = {Generative Agents: Interactive Simulacra of Human Behavior},
  year      = {2023},
  isbn      = {9798400701320},
  publisher = {Association for Computing Machinery},
  address   = {New York, NY, USA},
  url       = {https://doi.org/10.1145/3586183.3606763},
  doi       = {10.1145/3586183.3606763},
  booktitle = {Proceedings of the 36th Annual ACM Symposium on User Interface Software and Technology},
  articleno = {2},
  numpages  = {22},
  location  = {San Francisco, CA, USA},
  series    = {UIST '23}
}

@inproceedings{Chu2024,
  author    = {Chu, Zheng and Chen, Jingchang and Chen, Qianglong and Yu, Weijiang and He, Tao and Wang, Haotian and Peng, Weihua and Liu, Ming and Qin, Bing and Liu, Ting},
  title     = {{TimeBench}: A Comprehensive Evaluation of Temporal Reasoning Abilities in Large Language Models},
  year      = {2024},
  publisher = {Association for Computational Linguistics},
  url       = {https://aclanthology.org/2024.acl-long.66},
  doi       = {10.18653/v1/2024.acl-long.66},
  booktitle = {Proceedings of the 62nd Annual Meeting of the Association for Computational Linguistics (Volume 1: Long Papers)},
  pages     = {1204--1228}
}

@article{Pan2024,
  author    = {Pan, Shirui and Luo, Linhao and Wang, Yufei and Chen, Chen and Wang, Jiapu and Wu, Xindong},
  title     = {Unifying Large Language Models and Knowledge Graphs: A Roadmap},
  journal   = {IEEE Transactions on Knowledge and Data Engineering},
  volume    = {36},
  number    = {7},
  pages     = {3580--3599},
  year      = {2024},
  publisher = {IEEE},
  doi       = {10.1109/TKDE.2024.3352100},
  issn      = {1041-4347}
}

@misc{Edge2024,
  author        = {Edge, Darren and Trinh, Ha and Cheng, Newman and Bradley, Joshua and Chao, Alex and Mody, Apurva and Truitt, Steven and Larson, Jonathan},
  title         = {From Local to Global: A Graph {RAG} Approach to Query-Focused Summarization},
  year          = {2024},
  eprint        = {2404.16130},
  archivePrefix = {arXiv},
  primaryClass  = {cs.CL},
  url           = {https://arxiv.org/abs/2404.16130}
}

@inproceedings{Yasunaga2021,
  author    = {Yasunaga, Michihiro and Ren, Hongyu and Bosselut, Antoine and Liang, Percy and Leskovec, Jure},
  title     = {{QA-GNN}: Reasoning with Language Models and Knowledge Graphs for Question Answering},
  year      = {2021},
  publisher = {Association for Computational Linguistics},
  url       = {https://aclanthology.org/2021.naacl-main.45},
  doi       = {10.18653/v1/2021.naacl-main.45},
  booktitle = {Proceedings of the 2021 Conference of the North American Chapter of the Association for Computational Linguistics: Human Language Technologies},
  pages     = {535--546}
}

@inproceedings{Balog2019,
  author    = {Balog, Krisztian and Kenter, Tom},
  title     = {Personal Knowledge Graphs: A Research Agenda},
  year      = {2019},
  isbn      = {9781450368810},
  publisher = {Association for Computing Machinery},
  address   = {New York, NY, USA},
  url       = {https://doi.org/10.1145/3341981.3344241},
  doi       = {10.1145/3341981.3344241},
  booktitle = {Proceedings of the 2019 ACM SIGIR International Conference on Theory of Information Retrieval},
  pages     = {217--220},
  numpages  = {4},
  location  = {Santa Clara, CA, USA},
  series    = {ICTIR '19}
}

@inproceedings{Lewis2020,
  author    = {Lewis, Patrick and Perez, Ethan and Piktus, Aleksandra and Petroni, Fabio and Karpukhin, Vladimir and Goyal, Naman and K{\"u}ttler, Heinrich and Lewis, Mike and Yih, Wen-tau and Rockt{\"a}schel, Tim and Riedel, Sebastian and Kiela, Douwe},
  title     = {Retrieval-Augmented Generation for Knowledge-Intensive {NLP} Tasks},
  year      = {2020},
  publisher = {Curran Associates Inc.},
  address   = {Red Hook, NY, USA},
  booktitle = {Proceedings of the 34th International Conference on Neural Information Processing Systems},
  articleno = {793},
  numpages  = {16},
  pages     = {9459--9474},
  location  = {Vancouver, BC, Canada},
  series    = {NeurIPS '20}
}

@misc{Packer2023,
  author        = {Packer, Charles and Wooders, Sarah and Lin, Kevin and Fang, Vivian and Patil, Shishir G. and Stoica, Ion and Gonzalez, Joseph E.},
  title         = {{MemGPT}: Towards {LLMs} as Operating Systems},
  year          = {2023},
  eprint        = {2310.08560},
  archivePrefix = {arXiv},
  primaryClass  = {cs.AI},
  url           = {https://arxiv.org/abs/2310.08560}
}

@article{Dantzig2013,
  author    = {van Dantzig, S. and Geleijnse, G. and van Halteren, A. T.},
  title     = {Toward a persuasive mobile application to reduce sedentary behavior},
  journal   = {Personal and Ubiquitous Computing},
  volume    = {17},
  number    = {6},
  pages     = {1237--1246},
  year      = {2013},
  month     = {Aug},
  doi       = {10.1007/s00779-012-0588-0},
  url       = {https://doi.org/10.1007/s00779-012-0588-0}
}

@software{Goel2026,
    author = {Goel, Akshay},
    doi = {10.5281/zenodo.17015089},
    license = {Apache-2.0},
    month = may,
    title = {{LangExtract}},
    url = {https://github.com/google/langextract},
    version = {1.5.0},
    year = {2026}
}

@misc{Goel2025,
  author       = {Akshay Goel and Atilla Kiraly},
  title        = {Introducing {LangExtract}: A {Gemini} powered information extraction library},
  howpublished = {Google Developers Blog},
  month        = jul,
  year         = 2025,
  day          = 30,
  url          = {https://developers.googleblog.com/introducing-langextract-a-gemini-powered-information-extraction-library/},
  note         = {Accessed: [Insert access date here]}
}

@misc{Rosen2026,
      title={From Agent Loops to Deterministic Graphs: Execution Lineage for Reproducible AI-Native Work}, 
      author={Josh Rosen and Seth Rosen},
      year={2026},
      eprint={2605.06365},
      archivePrefix={arXiv},
      primaryClass={cs.AI},
      url={https://arxiv.org/abs/2605.06365}, 
}

@misc{Kim2025,
      title={Structured Cognitive Loop for Behavioral Intelligence in Large Language Model Agents}, 
      author={Myung Ho Kim},
      year={2025},
      eprint={2510.05107},
      archivePrefix={arXiv},
      primaryClass={cs.AI},
      url={https://arxiv.org/abs/2510.05107}, 
}

@article{Braun2006,
  author    = {Braun, Virginia and Clarke, Victoria},
  title     = {Using Thematic Analysis in Psychology},
  journal   = {Qualitative Research in Psychology},
  volume    = {3},
  number    = {2},
  pages     = {77--101},
  year      = {2006},
  publisher = {Routledge},
  doi       = {10.1191/1478088706qp063oa},
  issn      = {1478-0887}
}

@book{Boyatzis1998,
  author    = {Boyatzis, Richard E.},
  title     = {Transforming Qualitative Information: Thematic Analysis and Code Development},
  year      = {1998},
  publisher = {Sage Publications},
  address   = {Thousand Oaks, CA},
  isbn      = {9780761909613}
}

@book{Guest2012,
  author    = {Guest, Greg and MacQueen, Kathleen M. and Namey, Emily E.},
  title     = {Applied Thematic Analysis},
  year      = {2012},
  publisher = {Sage Publications},
  address   = {Thousand Oaks, CA},
  doi       = {10.4135/9781483384436},
  isbn      = {9781412971676}
}

@article{MacQueen1998,
  author    = {MacQueen, Kathleen M. and McLellan, Eleanor and Kay, Kelly and Milstein, Bobby},
  title     = {Codebook Development for Team-Based Qualitative Analysis},
  journal   = {CAM Journal},
  volume    = {10},
  number    = {2},
  pages     = {31--36},
  year      = {1998},
  publisher = {SAGE Publications},
  doi       = {10.1177/1525822X980100020301},
  issn      = {1525-822X}
}

@article{LandisKoch1977,
  author    = {Landis, J. Richard and Koch, Gary G.},
  title     = {The Measurement of Observer Agreement for Categorical Data},
  journal   = {Biometrics},
  volume    = {33},
  number    = {1},
  pages     = {159--174},
  year      = {1977},
  publisher = {JSTOR},
  doi       = {10.2307/2529310},
  issn      = {0006-341X}
}

@misc{Kim2026,
  title={Bridging Symbolic Control and Neural Reasoning in LLM Agents: Structured Cognitive Loop with a Governance Layer}, 
  author={Myung Ho Kim},
  year={2026},
  eprint={2511.17673},
  archivePrefix={arXiv},
  primaryClass={cs.AI},
  url={https://arxiv.org/abs/2511.17673}, 
}

@article{Chakraborty2023,
  title={A comprehensive survey of personal knowledge graphs},
  author={Chakraborty, Prantika and Sanyal, Debarshi Kumar},
  journal={Wiley Interdisciplinary Reviews: Data Mining and Knowledge Discovery},
  volume={13},
  number={6},
  pages={e1513},
  year={2023},
  publisher={Wiley Online Library}
}

@article{Hogan2021,
    author = {Hogan, Aidan and Blomqvist, Eva and Cochez, Michael and D’amato, Claudia and Melo, Gerard De and Gutierrez, Claudio and Kirrane, Sabrina and Gayo, Jos\'{e} Emilio Labra and Navigli, Roberto and Neumaier, Sebastian and Ngomo, Axel-Cyrille Ngonga and Polleres, Axel and Rashid, Sabbir M. and Rula, Anisa and Schmelzeisen, Lukas and Sequeda, Juan and Staab, Steffen and Zimmermann, Antoine},
    title = {Knowledge Graphs},
    year = {2021},
    issue_date = {May 2022},
    publisher = {Association for Computing Machinery},
    address = {New York, NY, USA},
    volume = {54},
    number = {4},
    issn = {0360-0300},
    url = {https://doi.org/10.1145/3447772},
    doi = {10.1145/3447772},
    journal = {ACM Comput. Surv.},
    month = jul,
    articleno = {71},
    numpages = {37},
}

@inproceedings{Zhu2010,
  title={Using finite state machines for evaluating spoken dialog systems},
  author={Zhu, Yi and Yang, Zhaojun and Meng, Helen and Li, Baichuan and Levow, Gina and King, Irwin},
  booktitle={2010 IEEE Spoken Language Technology Workshop},
  pages={478--483},
  year={2010},
  organization={IEEE}
}

@article{Young2013,
  title={Pomdp-based statistical spoken dialog systems: A review},
  author={Young, Steve and Ga{\v{s}}i{\'c}, Milica and Thomson, Blaise and Williams, Jason D},
  journal={Proceedings of the IEEE},
  volume={101},
  number={5},
  pages={1160--1179},
  year={2013},
  publisher={IEEE}
}

@misc{Zheng2023,
      title={Judging LLM-as-a-Judge with MT-Bench and Chatbot Arena}, 
      author={Lianmin Zheng and Wei-Lin Chiang and Ying Sheng and Siyuan Zhuang and Zhanghao Wu and Yonghao Zhuang and Zi Lin and Zhuohan Li and Dacheng Li and Eric P. Xing and Hao Zhang and Joseph E. Gonzalez and Ion Stoica},
      year={2023},
      eprint={2306.05685},
      archivePrefix={arXiv},
      primaryClass={cs.CL},
      url={https://arxiv.org/abs/2306.05685}, 
}

@misc{Openai2023,
      title={GPT-4 Technical Report}, 
      author={OpenAI et al.},
      year={2023},
      eprint={2303.08774},
      archivePrefix={arXiv},
      primaryClass={cs.CL},
      url={https://arxiv.org/abs/2303.08774}, 
}

@article{Allan1983,
author = {Allen, James F.},
title = {Maintaining knowledge about temporal intervals},
year = {1983},
issue_date = {Nov. 1983},
publisher = {Association for Computing Machinery},
address = {New York, NY, USA},
volume = {26},
number = {11},
issn = {0001-0782},
url = {https://doi.org/10.1145/182.358434},
doi = {10.1145/182.358434},
journal = {Commun. ACM},
month = nov,
pages = {832–843},
numpages = {12}
}

@misc{Menschikov2026,
      title={PersonalAI: A Systematic Comparison of Knowledge Graph Storage and Retrieval Approaches for Personalized LLM agents}, 
      author={Mikhail Menschikov and Dmitry Evseev and Victoria Dochkina and Ruslan Kostoev and Ilia Perepechkin and Petr Anokhin and Nikita Semenov and Evgeny Burnaev},
      year={2026},
      eprint={2506.17001},
      archivePrefix={arXiv},
      primaryClass={cs.CL},
      url={https://arxiv.org/abs/2506.17001}, 
}

@inproceedings{Wang2024,
  title     = {{TRAM}: Benchmarking Temporal Reasoning for Large Language Models},
  author    = {Wang, Yuqing and Zhao, Yun},
  booktitle = {Findings of the Association for Computational Linguistics: ACL 2024},
  month     = aug,
  year      = {2024},
  address   = {Bangkok, Thailand},
  publisher = {Association for Computational Linguistics},
  url       = {https://aclanthology.org/2024.findings-acl.382/},
  doi       = {10.18653/v1/2024.findings-acl.382},
  pages     = {6389--6415}
}

@article{Maharjan2022,
  author    = {Maharjan, Raju and Doherty, Kevin and Rohani, Darius Adam and B{\ae}kgaard, Per and Bardram, Jakob E.},
  title     = {Experiences of a Speech-Enabled Conversational Agent for the Self-Report of Well-Being Among People Living with Affective Disorders: An In-the-Wild Study},
  journal   = {ACM Transactions on Interactive Intelligent Systems},
  year      = {2022},
  publisher = {Association for Computing Machinery},
  doi       = {10.1145/3484508},
  issn      = {2160-6455}
}

@inproceedings{Caine2016,
author = {Caine, Kelly},
title = {Local Standards for Sample Size at CHI},
year = {2016},
isbn = {9781450333627},
publisher = {Association for Computing Machinery},
address = {New York, NY, USA},
url = {https://doi.org/10.1145/2858036.2858498},
doi = {10.1145/2858036.2858498},
booktitle = {Proceedings of the 2016 CHI Conference on Human Factors in Computing Systems},
pages = {981–992},
numpages = {12},
location = {San Jose, California, USA},
series = {CHI '16}
}

@incollection{Clark1991,
  title={Grounding in communication},
  author={Clark, Herbert H. and Brennan, Susan E.},
  booktitle={Perspectives on socially shared cognition},
  editor={Resnick, L. B. and Levine, J. M. and Teasley, S. D.},
  pages={127--149},
  year={1991},
  publisher={American Psychological Association},
  doi={10.1037/10096-006}
}

@article{Pinder2018,
author = {Pinder, Charlie and Vermeulen, Jo and Cowan, Benjamin R. and Beale, Russell},
title = {Digital Behaviour Change Interventions to Break and Form Habits},
year = {2018},
issue_date = {June 2018},
publisher = {Association for Computing Machinery},
address = {New York, NY, USA},
volume = {25},
number = {3},
issn = {1073-0516},
url = {https://doi.org/10.1145/3196830},
doi = {10.1145/3196830},
journal = {ACM Trans. Comput.-Hum. Interact.},
month = jun,
articleno = {15},
numpages = {66}
}

@article{Berube2024,
  title={Proactive behavior in voice assistants: A systematic review and conceptual model},
  author={B{\'e}rub{\'e}, Caterina and Ni{\ss}en, Marcia and Vinay, Rasita and Geiger, Alexa and Budig, Tobias and Bhandari, Aashish and Pe Benito, Catherine Rachel and Ibarcena, Nathan and Pistolese, Olivia and Li, Pan and Bin Sawad, Abdullah and Fleisch, Elgar and Stettler, Christoph and Hemsley, Bronwyn and Berkovsky, Shlomo and Kowatsch, Tobias and Kocaballi, A. Baki},
  journal={Computers in Human Behavior Reports},
  volume={14},
  pages={100411},
  year={2024},
  issn={2451-9588},
  doi={10.1016/j.chbr.2024.100411},
  url={https://doi.org/10.1016/j.chbr.2024.100411}
}

@misc{Salwa2025,
  title={Continual Learning: Overcoming Catastrophic Forgetting for Adaptive AI Systems},
  author={Salwa, Husniya and Burhan, Ntombifuthi and Rahel, Ernest},
  year={2025},
  month={Feb},
  publisher={TechRxiv},
  doi={10.36227/techrxiv.173886426.63028528/v1},
  url={https://doi.org/10.36227/techrxiv.173886426.63028528/v1}
}

@article{Planer2023,
  title={The evolution of hierarchically structured communication},
  author={Planer, R. J.},
  journal={Frontiers in Psychology},
  volume={14},
  pages={1224324},
  year={2023},
  doi={10.3389/fpsyg.2023.1224324},
  url={https://doi.org/10.3389/fpsyg.2023.1224324}
}

@inproceedings{ReinhartzBerger2025,
  author    = {Reinhartz-Berger, I. and Ali, S. J. and Bork, D.},
  editor    = {Krogstie, J. and Rinderle-Ma, S. and Kappel, G. and Proper, H. A.},
  title     = {Leveraging LLMs for Domain Modeling: The Impact of Granularity and Strategy on Quality},
  booktitle = {Advanced Information Systems Engineering. CAiSE 2025},
  series    = {Lecture Notes in Computer Science},
  volume    = {15701},
  publisher = {Springer, Cham},
  year      = {2025},
  doi       = {10.1007/978-3-031-94569-4_1},
  url       = {https://doi.org/10.1007/978-3-031-94569-4_1}
}

@article{Farrahi2011,
  author = {Farrahi, Katayoun and Gatica-Perez, Daniel},
  title = {Discovering routines from large-scale human locations using probabilistic topic models},
  journal = {ACM Transactions on Intelligent Systems and Technology (TIST)},
  volume = {2},
  number = {1},
  pages = {1--27},
  year = {2011},
  publisher = {ACM New York, NY, USA},
  doi = {10.1145/1889681.1889684}
}

@article{Zacks2001,
  title={Event structure in perception and conception},
  author={Zacks, Jeffrey M. and Tversky, Barbara},
  journal={Psychological Bulletin},
  volume={127},
  number={1},
  pages={3--21},
  year={2001},
  publisher={American Psychological Association},
  doi={10.1037/0033-2909.127.1.3}
}

@article{CanabalJuanatey2024,
  title={Enriching interactive explanations with fuzzy temporal constraint networks},
  author={Canabal-Juanatey, Mari{\~n}a and Alonso-Moral, Jose M. and Catala, Alejandro and Bugar{\'\i}n-Diz, Alberto},
  journal={International Journal of Approximate Reasoning},
  volume={171},
  pages={109128},
  year={2024},
  issn={0888-613X},
  doi={10.1016/j.ijar.2024.109128},
  url={https://doi.org/10.1016/j.ijar.2024.109128}
}

@article{Dwivedi2023,
author = {Dwivedi, Rudresh and Dave, Devam and Naik, Het and Singhal, Smiti and Omer, Rana and Patel, Pankesh and Qian, Bin and Wen, Zhenyu and Shah, Tejal and Morgan, Graham and Ranjan, Rajiv},
title = {Explainable AI (XAI): Core Ideas, Techniques, and Solutions},
year = {2023},
issue_date = {September 2023},
publisher = {Association for Computing Machinery},
address = {New York, NY, USA},
volume = {55},
number = {9},
issn = {0360-0300},
url = {https://doi.org/10.1145/3561048},
doi = {10.1145/3561048},
journal = {ACM Comput. Surv.},
month = jan,
articleno = {194},
numpages = {33}
}

@inproceedings{Xu2019,
  author    = {Xu, F. and Uszkoreit, H. and Du, Y. and Fan, W. and Zhao, D. and Zhu, J.},
  editor    = {Tang, J. and Kan, M. Y. and Zhao, D. and Li, S. and Zan, H.},
  title     = {Explainable AI: A Brief Survey on History, Research Areas, Approaches and Challenges},
  booktitle = {Natural Language Processing and Chinese Computing. NLPCC 2019},
  series    = {Lecture Notes in Computer Science},
  volume    = {11839},
  publisher = {Springer, Cham},
  year      = {2019},
  doi       = {10.1007/978-3-030-32236-6_51},
  url       = {https://doi.org/10.1007/978-3-030-32236-6_51}
}

@inproceedings{Kulesza2015,
author = {Kulesza, Todd and Burnett, Margaret and Wong, Weng-Keen and Stumpf, Simone},
title = {Principles of Explanatory Debugging to Personalize Interactive Machine Learning},
year = {2015},
isbn = {9781450333061},
publisher = {Association for Computing Machinery},
address = {New York, NY, USA},
url = {https://doi.org/10.1145/2678025.2701399},
doi = {10.1145/2678025.2701399},
pages = {126–137},
numpages = {12},
location = {Atlanta, Georgia, USA},
series = {IUI '15}
}

@inproceedings{Liao2020,
author = {Liao, Q. Vera and Gruen, Daniel and Miller, Sarah},
title = {Questioning the AI: Informing Design Practices for Explainable AI User Experiences},
year = {2020},
isbn = {9781450367080},
publisher = {Association for Computing Machinery},
address = {New York, NY, USA},
url = {https://doi.org/10.1145/3313831.3376590},
doi = {10.1145/3313831.3376590},
booktitle = {Proceedings of the 2020 CHI Conference on Human Factors in Computing Systems},
pages = {1–15},
numpages = {15},
location = {Honolulu, HI, USA},
series = {CHI '20}
}

\appendix

\end{document}